\documentclass[11pt]{iopart}
\usepackage{iopams}  
\usepackage{graphicx}
\usepackage{amssymb}

\begin{document}

\title[Measurement of Electron Sheath Thickness and Collection Region]{Measurement of Electron Sheath Thickness and Collection Region of Electric Probe using Laser Photodetachment Signals}

\author{S. Kajita\dag
\footnote[1]{To
E-mail address: shin@flanker.q.t.u-tokyo.ac.jp}
 S. Kado\ddag\
\footnote[2]{To
E-mail address: kado@q.t.u-tokyo.ac.jp}
A. Okamoto\ddag\, T. Shikama\dag\, Y. Iida\dag\, D. Yamasaki\dag\, and S. Tanaka\dag\
}

\address{\dag\ Department of Quantum Engineering and Systems Science,
Graduate School of Engineering, The University of Tokyo, 7-3-1 Hongo, Bunkyo, Tokyo 113-8656 Japan}

\address{\ddag\ High Temperature Plasma Center, The University of Tokyo, 2-11-16 Yayoi, Bunkyo, Tokyo 113-8656 Japan}
\begin{abstract}
A new type of laser photodetachment (LPD) technique has been developed for the measurement of electron sheath thickness around an electrostatic probe and for the measurement of the length of collection region of photodetached electrons (PDE).
When a thickness of the sheath formed around an electrostatic probe is thicker than about 0.1 mm, modification of the temporal evolution of the LPD signal is observed. By making use of the modification, we evaluated sheath thickness around a cylindrical probe in the existence of the magnetic field of 15 mT. It was found that the thickness of electron sheath along the magnetic field was comparable to the calculated plane-parallel Child-Langmuir sheath thickness when the probe-bias voltage was high. Furthermore, by inserting a small screening object in the laser beam channel, we can measure the sheath thickness from the modification of LPD signal. From the variation of the signal intensity as scanning the screening object perpendicular to the laser beam channel, we can observe the collection region of photodetached electrons, because this procedure changes the relative displacement between shadow and the probe electrode.
\end{abstract}




\section{Introduction}
The laser photodetachment (LPD) technique combined with an electrostatic probe has been widely applied to the measurements of negative ion density and the drift velocity \cite{Bacal3}. In this measurement, the excess electrons, produced by the laser-induced photodetachment process, ${\rm H^-}+h\nu \to {\rm H+e^-}$, are detected by a positively biased electrostatic probe. Then, we can determine the negative ion density $n_{-}$ and the drift velocity $v_{\rm d}$ on the basis of the peak value of the excess electron current $\Delta I_{\rm e}$ and of the temporal evolution of the electron current response, respectively.

Recently, it has been pointed out that negative ions play an important role in the divertor region for experimental fusion reactors. The negative ions are expected to contribute to the enhancement of recombination processes, which are capable of reducing the heat flux to the divertor plate \cite{Janev,Krash}. Authors have been applied LPD technique to divertor simulator MAP (Material And Plasma) -II for the purpose of clarifying the contribution of negative ions to the plasma recombination \cite{Kajita1, Kado2}.
Up to now, however, the applicability of the laser photodetachment technique under the existence of magnetic field has not been examined. In particular, {\it in situ} measurement of not only the sheath around an electrostatic probe but also of the collection region of photodetached electrons (PDE) stretches to the outside of the sheath is desirable for this purpose.
In the present paper, the developed new measurements of electron sheath thickness and the length of the collection region of PDE are presented.

\section{Principle of the measurement}
\subsection{Measurement of electron sheath thickness}
The recovery time of LPD signal $t_{\rm recov}$, in which the electron density at the center of the laser beam recovers to the initial density, is expressed using the drift velocity of negative ions $v_{\rm d}$ and the radius of laser beam $R_{\rm L}$ \cite{Stern}. The recovery time of the LPD signal is expressed as
\begin{eqnarray}
t_{\rm recov} \simeq \frac{R_{\rm L}-(r_{\rm p}+h)}{v_{\rm d}}, \label{eq1-2}
\end{eqnarray}
where $r_{\rm p}$ represents the radius of the cylindrical probe tip and $h$ the thickness of the electron sheath. Because the sheath thickness is negligible at the space potential, the recovery time at the space potential $t_{\rm recov0}$ is deduced by substituting $h=0$ in Eq. (\ref{eq1-2}), as follows:
\begin{eqnarray}
t_{\rm recov0} \simeq \frac{R_{\rm L}-r_{\rm p}}{v_{\rm d}}. \label{tr_modif}
\end{eqnarray}
Thus, the sheath thickness is obtained from Eqs. (\ref{eq1-2}) and (\ref{tr_modif}), as
\begin{eqnarray}
h \simeq \frac{(t_{\rm recov0}-t_{\rm recov})(R_{\rm L}-r_{\rm p})}{t_{\rm recov 0}}. \label{sheath1}
\end{eqnarray}

\begin{figure}[tb]
\begin{center}
\includegraphics[width=0.75\linewidth]{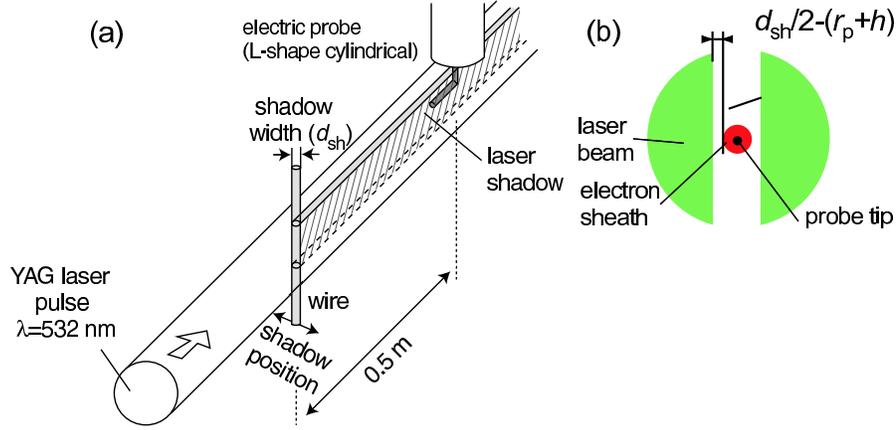}
\caption{(a) Schematic view of experimental setup for eclipse laser photodetachment. (b) The configuration of laser beam channel and probe tip in the eclipse-LPD method.}
\label{Fig:eclipse}
\end{center}
\end{figure}

Next, let us consider the case that a screening object is inserted in the laser beam channel. Originally, the screening object is used to avoid the probe surface ablation \cite{KajitaContrib} by shielding the probe tip from the direct laser irradiation \cite{BacalScreen}. We named this screening method ``eclipse laser photodetachment method'' after a lunar eclipse, in which the shadow of the earth protects the moon from irradiation by the sun. It was confirmed that this eclipse-LPD method provides the proper negative ion density when the screening object is sufficiently smaller than the laser size \cite{KajitaEPS, KajitaPSS}.
A schematic view of the experimental arrangement of the eclipse-LPD is depicted in Fig. \ref{Fig:eclipse} (a). A thin wire was inserted in the laser beam channel to form a thin shadow in the laser channel. The position of the shadow was changed with a micrometer of the wire stage. Note that the width of the shadow becomes a little wider than the screening object because of the diffraction of the laser beam. Taking into consideration the distance from the wire to the probe tip ($\sim$0.5 m) and the wavelength of the laser (532 nm), the calculation based on the Fresnel diffraction provides the effective shadow width $d_{\rm sh}$ of the wire diameter plus 0.6 mm \cite{KajitaPSS}.

\begin{figure}[tbp]
\begin{center}
\includegraphics[width=0.5\linewidth]{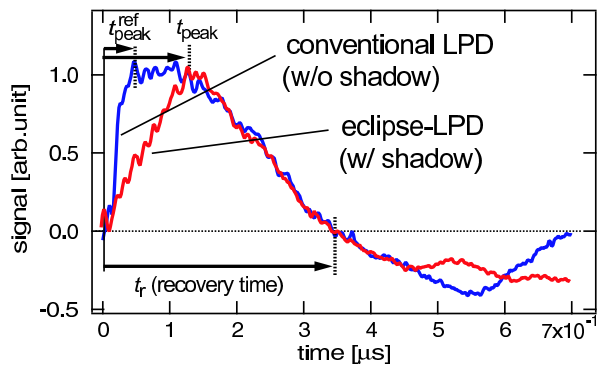}
\caption{Typical conventional laser photodetachment signal and eclipse laser photodetachment signal. ($V_{\rm p}-V_{\rm sp}=30$ V)}
\label{Fig:Signals}
\end{center}
\end{figure}

Typical temporal evolution of the conventional LPD signal and that of the eclipse-LPD are shown in Fig. \ref{Fig:Signals}. We have checked that the recovery time is not disturbed by the screening object when the shadow width is sufficiently thinner than the laser diameter. Thus the sheath effects can be observed in both eclipse-LPD signals and in conventional LPD signals as the shift of the recovery time.

On the other hand, the time for the LPD signal to reach its peak value depends on the displacement between the effective probe surface including the sheath and the edge of the photodetached electron {\it swarm}. The configuration of the laser beam and probe tip in eclipse-LPD are shown in Fig. \ref{Fig:eclipse} (b). 
When the probe is located outside the laser-irradiated area as shown in Fig. \ref{Fig:eclipse} (b), the peak time of the signal $t_{\rm peak}$ is shifted from $t_{\rm peak}^{\rm ref}$ by $\Delta t_{\rm peak}$.
Because the propagation velocity of the excess electron {\it swarm} toward the outside of the laser irradiated region corresponds to the negative ion drift velocity \cite{Stern}, the time shift due to the propagation can be expressed according to Fig. \ref{Fig:eclipse} (b) as 
\begin{eqnarray}
\Delta t_{\rm peak}(h)=t_{\rm peak}(h) -t_{\rm peak}^{\rm ref} \simeq \frac{d_{\rm sh}/2-(r_{\rm p}+h)}{v_{\rm d}} \hspace{0.5cm}(>0),\label{ptim_modif}
\end{eqnarray}
where $d_{\rm sh}$ is the shadow width and the term $(r_{\rm p} +h)$ represents the probe radius corrected by the sheath thickness.
The shift of the peak time at the space potential $\Delta t_{\rm peak0}$ can be also deduced by substituting $h=0$ in Eq. (\ref{ptim_modif}) as
\begin{eqnarray}
\Delta t_{\rm peak0} = t_{\rm peak}(h=0) -t_{\rm peak}^{\rm ref} \simeq \frac{d_{\rm sh}/2-r_{\rm p}}{v_{\rm d}}. \label{ptim_modif0}
\end{eqnarray}
Finally, the electron sheath thickness is deduced from the difference between Eqs. (\ref{ptim_modif}) and (\ref{ptim_modif0}) as
\begin{eqnarray}
h \simeq (\Delta t_{\rm peak0}-\Delta t_{\rm peak})v_{\rm d}, \label{sheath2}
\end{eqnarray}
where $v_{\rm d}$ can be obtained from Eq. (\ref{tr_modif}). 
Moreover, from Eqs. (\ref{eq1-2}) and (\ref{ptim_modif}), the sheath thickness is obtained as
\begin{eqnarray}
h \simeq \frac{(d_{\rm sh}/2-r_{\rm p})t_{\rm recov}-(R_{\rm L}-r_{\rm p})\Delta t_{\rm peak}}{\Delta t_{\rm peak}+t_{\rm recov}}. \label{sheath3}
\end{eqnarray}
Therefore, in the eclipse-LPD method, we can obtain the sheath thickness $h$ in three independent ways. However, use of either Eq. (\ref{sheath1}) or of Eqs. (\ref{sheath2}) and (\ref{tr_modif}) requires two signals at different bias voltages, space potential, and positive bias. In contrast, there is a significant merit in the application of Eq. (\ref{sheath3}) because sheath thickness can be measured from a single signal at any bias voltage.

\subsection{Measurement of collection region}
As mentioned above, the time evolution of the LPD signal depends on the electron sheath thickness. On the other hand, the intensity of the LPD signal is sensitive to the dimension of the collection region of PDE.
In the eclipse-LPD method, the signal intensity decreases when part of the collection region is shaded by the shadow. However, the signal intensity recovers when the shadow is well aligned with the electrostatic probe.
This phenomena can be explained by the excess electrons leaked from the irradiated region to the shaded region \cite{KajitaPSS}. The width of the total signal dip is $2 (L_{\rm PDE} +  r_{\rm p} + d_{\rm sh}/2 )$, where $L_{\rm PDE}$ is the length of the collection region of PDE. We can thus obtain $L_{\rm PDE}$.

\section{Experiments}
The experiments were performed in divertor simulator MAP-II, which consists of differentially pumped dual chambers, target chamber and source chamber. The chambers are connected with a drift tube. The plasma is generated between a LaB$_6$ cathode and an anode pipe by an arc discharge.
The magnetic field of 15 mT is formed with 8 solenoid coils, and then, a cylindrical plasma of about 2 m in length is formed.
The electron density $n_{\rm e}$ and the temperature $T_{\rm e}$ at the center of the plasma column in the target chamber are about $10^{12}$ cm$^{-3}$ and $5$ eV, respectively, while those at the peripheral region ($r\sim$5 cm) are about $10^{11}$ cm$^{-3}$ and $1$ eV, respectively.
An L-shaped electrostatic probe, consists of a 0.3 mm in diameter tungsten wire, was placed on the axis perpendicular to the magnetic field in the target chamber. The total length exposed to the plasma was 5 mm (2mm plus 3 mm). The laser was injected parallel to the probe tip head.
In the present paper, the probe was located at the peripheral region of the plasma column, where the negative ion density ratio to the electron density is about several percent \cite{Kajita1, Kado2}.
The negative ion density ratio was so small that we neglected the modification of the sheath structure due to the photodetachment process. 
Second harmonic Nd:YAG laser pulses (the wavelength of $\lambda=$532 nm) were used for the photon source of the photodetachment. A thin wire for the eclipse-LPD is located outside the vacuum chamber. The distance from the wire to the probe tip was about 0.5 m and the shadow position was adjusted with a micrometer of the wire stage.
In the case of 15 mT, the Larmor radius of the electrons is comparable to the size of the probe. In order to examine the effect of the magnetic field, the case without the magnetic field was compared with the case at 15 mT. 
In reality, a weak magnetic field of less than 1 mT exists even when the coil currents at the target chamber are turned off. However, the effect is negligible because the Larmor radius of electrons is much larger than the probe size in this case. Thus, we henceforth call this condition 0 mT.

\section{Results and discussion}
\begin{figure}[tbp]
\begin{center}
\includegraphics[width=0.5\linewidth]{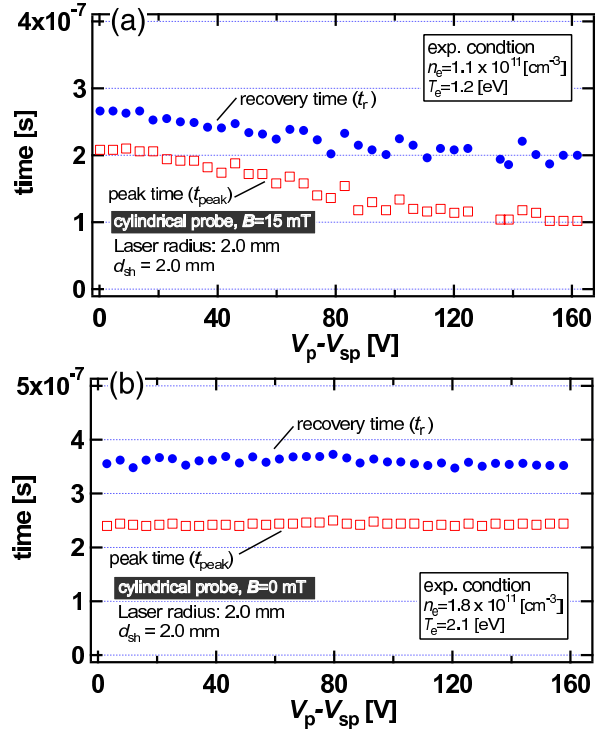}
\caption{The recovery time and the peak time of LPD signal as a function of probe-bias voltage. (a) is the case of 15 mT  (b) is the case of 0 mT [shot $\sharp$13926, $\sharp$14393].}
\label{Fig:Sheath}
\end{center}
\end{figure}

The recovery time and the peak time of eclipse-LPD signals as a function of the probe-bias voltage are shonw in Fig. \ref{Fig:Sheath}. Figure \ref{Fig:Sheath} (a) and (b) correspond to the cases at 15 mT and at 0 mT, respectively. 
Both $t_{\rm recov}$ and $t_{\rm peak}$ were significantly reduced as increasing the probe-bias voltage in Fig. \ref{Fig:Sheath} (a). These reductions in Fig. \ref{Fig:Sheath} (a) attributes to the expansion of sheath thickness with the probe voltage.

On the other hand, the time shifts were not observed on either $t_{\rm recov}$ or $t_{\rm peak}$ in Fig. \ref{Fig:Sheath} (b).
The comparison between Fig. \ref{Fig:Sheath} (a) and (b) shows that the thickness of sheath around the cylindrical probe at 0 mT is much thinner than that at 15 mT. 
In this paper, we focus to the case of the cylindrical probe at 15 mT, because the modifications of $t_{\rm recov}$ and $t_{\rm peak}$ at 0 mT were so small that it was difficult to deduce the sheath thickness.

For the purpose of comparing them to the experimental results with theory, the sheath thickness was estimated using the Child-Langmuir law assuming that the particle energy is sufficiently small compared to the potential of the sheath namely, $T_{\rm e} \ll e(V_{\rm p}-V_{\rm sp})$, where $T_{\rm e}$ is the electron temperature in eV, $e$ the elementary charge, $V_{\rm p}$ the probe voltage, and $V_{\rm sp}$ the space potential. The electron sheath thickness in one-dimensional geometry can be written as \cite{Child-Langmuir}
\begin{eqnarray}
h_{\rm p}=\frac{2}{3}\sqrt{-\frac{\epsilon_0}{j}}\left( \frac{2e(V_{\rm p}-V_{\rm sp})^3}{m_{\rm e}} \right)^{1/4}, \label{Sheath:PL}
\end{eqnarray}
where $j$ is the current density, $\epsilon_0$ the dielectric constant in vacuum, $m_{\rm e}$ the mass of an electron. In the present paper, $h_{\rm p}$ stands for the thickness of the ``plane-parallel Child-Langmuir (C-L) sheath''. It has been confirmed in MAP-II that the experimentally obtained thicknesses of the sheath around an plane probe are well agree with $h_{\rm p}$ in magnetized plasmas \cite{KajitaPRE}.
On the other hand, in cylindrical geometry, the sheath thickness is expressed as a function of $\beta=(r_{\rm p}+h)/r_{\rm p}$, where $r_{\rm p}$ is the probe radius, and then can be written as \cite{Langmuir}
\begin{eqnarray}
h_{\rm c}=-\frac{8\pi \epsilon_0}{9 j \beta^2}\sqrt{ \frac{2e}{m_e}}(V_{\rm p}-V_{\rm sp})^{3/2}. \label{Sheath:CY}
\end{eqnarray}
$h_{\rm c}$ stands for the thickness of the ``cylindrical Child-Langmuir (C-L) sheath'' in this paper. This can be regarded as the limit of using a cylindrical probe in un-magnetized plasmas. The cylindrical C-L sheath is much thinner than the plane parallel C-L sheath as is shown next.

\begin{figure}[tbp]
\begin{center}
\includegraphics[width=0.7\linewidth]{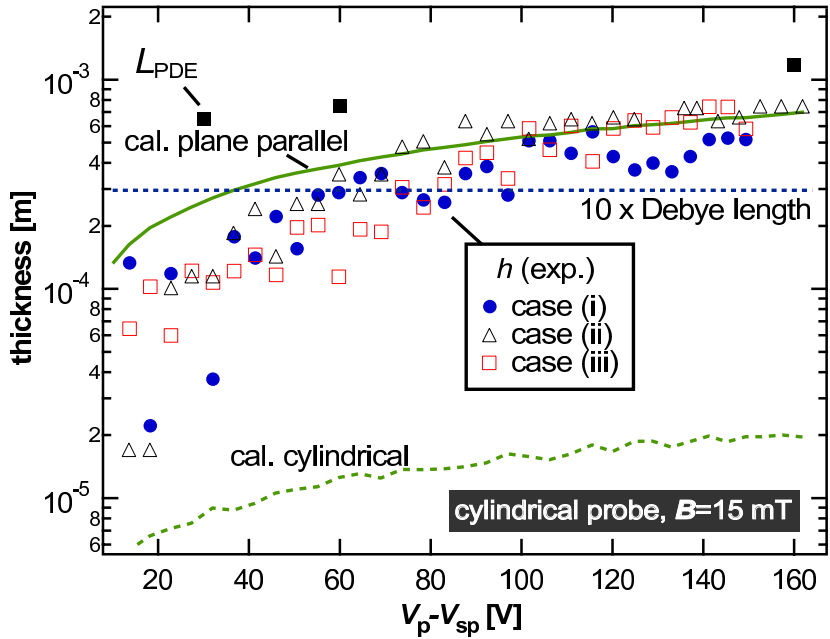}
\caption{Comparisons of experimentally obtained thickness of the electron sheath around a cylindrical probe at 15 mT, deduced from three different procedures (i), (ii) and (iii), to the calculated thicknesses of plane-parallel C-L sheath and cylindrical C-L sheath. Ten times the Debye length is also plotted as a dotted line. ($n_{\rm e}=1 \times 10^{11} {\rm cm^{-3}}$, $T_{\rm e}=1.6$ eV.) [shot $\sharp$13926].}
\label{Fig:Sheath_cyl}
\end{center}
\end{figure}

The electron sheath thicknesses obtained (i) from Eq. (\ref{sheath1}), (ii) from Eqs. (\ref{sheath2}) and (\ref{tr_modif}), and (iii) from Eq. (\ref{sheath3}), respectively, are plotted in Fig. \ref{Fig:Sheath_cyl}. The sheath thicknesses obtained from different procedures (i), (ii) and (iii) are consistent value with each other. In Fig. \ref{Fig:Sheath_cyl}, the calculated sheath thicknesses $h_{\rm p}$ and $h_{\rm c}$ are also plotted. The experimental results are much thicker than the cylindrical C-L sheath thickness, and moreover, they are agree with the calculated plane-parallel C-L sheath thickness at the probe-bias voltage higher than about 50 V. 
This result indicates that the effect of the magnetic field on the sheath structure is significant even in a weakly magnetized regime for electrons, where the Larmor radius of electron is comparable to the size of the probe tip.

In addition, $L_{\rm PDE}$ is measured at three different probe-bias voltages as changing the shadow position, and the results are also plotted in Fig. \ref{Fig:Sheath_cyl}. The length $L_{\rm LPD}$ also increases with the probe voltage, and is about three to five times thicker than $h$. Ten times the Debye length $\lambda_{\rm D}$, which is defined as $\lambda_{\rm D}=(\epsilon_0 T_{\rm e}/n_{\rm e}e^2)^{1/2}$, is also plotted in Fig. \ref{Fig:Sheath_cyl}, since $5\lambda_{\rm D}-10\lambda_{\rm D}$ has been regarded as the typical sheath thickness conventionally. When the probe was positively 70 - 80 V biased against the space potential, the sheath thickness was comparable to ten times the Debye length.

We can check the applicability of the negative ion velocity measurement with $h$, and that of the negative ion density measurement with $L_{\rm PDE}$. If a thick electron sheath is formed around an electrostatic probe, we should modify the recovery time by replacing $r_{\rm p}$ with $r_{\rm p} + h$. If $L_{\rm PDE}+r_{\rm p}$ is longer than the laser radius $R_{\rm L}$, we should enlarge the laser size or reducing the probe-bias voltage until the condition $L_{\rm PDE}+r_{\rm p}<R_{\rm L}$ is satisfied. Because $L_{\rm PDE}$ might be elongated along the magnetic field, it is strongly recommended to check the condition $L_{\rm PDE}+r_{\rm p}<R_{\rm L}$ especially when the LPD technique is applied to strongly magnetized plasmas. 

\section{Conclusions}
Electron sheath thicknesses around a cylindrical probe in weakly magnetized plasmas were measured at different probe-bias voltages by using a newly developed eclipse-LPD. The sheath thickness was consistent with the calculated sheath thickness from the plane-parallel Child-Langmuir law, especially when the probe voltage was rather high. In addition, the electron collection region of photodetached electrons was measured from the shadow position dependence of LPD signal intensity. Under the experimental condition in the present paper, the length of the electron collection region was three to five times longer than the electron sheath thickness.
The eclipse-LPD has been shown to be having in itself the capability of checking the applicability.


\section*{Reference}
\begin{thebibliography}{99} 
\bibitem{Bacal3} Bacal M 2000 {\it Rev. Sci. Instrum.} \textbf{71} 3981
\bibitem{Janev} Janev R K, Post D E, Langer W D, Evans K, Heifetz D B and Weisheit J C 1984 {\it J. Nucl. Mater.} \textbf{121} 10
\bibitem{Krash} Krasheninnikov I, Pigarov A Yu and Sigmar D J 1996 {\it Phys. Lett. A} \textbf{214} 295
\bibitem{Kajita1} Kajita S, Kado S, Uchida N, Shikama T and Tanaka S 2003 {\it J. Nucl. Mater.} \textbf{313-316}, 748
\bibitem{Kado2} Kado S, Kajita S, Yamasaki D, Iida Y, Xiao B, Shikama T, Oishi T, Okamoto A and Tanaka S, "Negative Ion Profiles in H$_2$-MAR Plasmas in Divertor Simulator MAP-II", to be published in {\it J. Nucl. Mater.}
\bibitem{Stern} Stern R A, Devynck P, Bacal M, Berlemont P and Hillion F 1990 {\it Phys. Rev. A} \textbf{41} 3307
\bibitem{KajitaContrib} Kajita S, Kado S, Shikama T, Xiao B and Tanaka S 2004 {\it Contrib. Plasma Phys.} \textbf{44} 607 (in press)
\bibitem{BacalScreen} Bacal M, Bruneteau A M and Nachman M 1981 {\it J. Phys. Lett.} \textbf{42} L-5
\bibitem{KajitaEPS} Kajita S, Kado S, {\it et al} 2004 {\it Proc. of the 31st EPS Conf. on plasma phys.}
\bibitem{KajitaPSS} Kajita S, Kado S and Tanaka S, "Eclipse laser photodetachment method to avoid probe surface ablation in negative ion measurement", submitted to {\it Plasma Sources Sci. and Technol.}
\bibitem{Child-Langmuir} Raizer Y P {\it Gas Discharge Physics} (1991 Springer-Verlag)
\bibitem{KajitaPRE} Kajita S, Kado S, Okamoto A and Tanaka S, "Application of Eclipse Laser Photodetachment Technique to Electron Sheath Thickness and Collection Region Measurements", to be published in {\it Phys. Rev. E}
\bibitem{Langmuir} Langmuir I and Blodgett K B 1925 {\it Phys. Rev.} \textbf{22} 347
\end{thebibliography}
\end{document}